\documentclass[12pt,preprint]{aastex}
\usepackage{natbib}
\usepackage{graphicx}
\usepackage{amsmath}
\usepackage{epsfig}
\begin{document}
   \title{Broad N$_2$H$^{+}$ emission towards the protostellar shock L1157-B1}

   \author{C. Codella$^1$, S. Viti$^2$, C. Ceccarelli$^3$, B. Lefloch$^3$, M. Benedettini$^4$, G. Busquet$^4$, P. Caselli$^5$, F. Fontani$^1$, A. G\'omez-Ruiz$^1$, L. Podio$^3$, M. Vasta$^1$}

\altaffiltext{1}{INAF, Osservatorio Astrofisico di Arcetri, Largo
Enrico Fermi 5, I-50125 Firenze, Italy \\ codella@rcetri.astro.it}
\altaffiltext{2}{Department of Physics and Astronomy, University
College London, London, UK}
\altaffiltext{3}{UJF-Grenoble 1 / CNRS-INSU, Institut de
Plan\'etologie et d'Astrophysique de Grenoble (IPAG) UMR 5274,
Grenoble, F-38041, France}
\altaffiltext{4}{INAF, Istituto di Astrofisica e Planetologia Spaziali,
via Fosso del Cavaliere 100, 00133, Roma, Italy}
\altaffiltext{5}{School of Physics and Astronomy, University of Leeds,
Leeds LS2 9JT, UK}

\date{Received - ; accepted -}
\begin{abstract}
We present the first detection of N$_2$H$^+$ towards
a low-mass protostellar outflow, namely the L1157-B1 shock, at $\sim$ 0.1 pc from the protostellar cocoon.   
The detection was obtained with the IRAM 30-m antenna. We observed
emission at 93 GHz due to the $J$ = 1--0 hyperfine lines.  
The analysis of the emission coupled with the HIFI CHESS multiline CO observations leads to
the conclusion that the observed N$_2$H$^+$(1--0) line  
originates from the dense ($\ge$ 10$^5$ cm$^{-3}$) gas associated with the 
large (20$\arcsec$--25$\arcsec$) cavities opened by the protostellar wind. 
We find a N$_2$H$^+$ column density of few 10$^{12}$ cm$^{-2}$  
corresponding to an abundance of (2--8) $\times$ 10$^{-9}$.  
The N$_2$H$^+$ abundance can be matched by a model of quiescent gas 
evolved for more than 10$^4$ yr, i.e. for 
more than the shock kinematical age ($\simeq$ 2000 yr).
Modelling of C-shocks confirms that the abundance of N$_2$H$^+$ is not
increased by the passage of the shock. 
In summary, N$_2$H$^+$ is a fossil record of the pre-shock gas, formed 
when the density of the gas was around 10$^4$ cm$^{-3}$, and then
further compressed and accelerated by the shock. 
\end{abstract}

\keywords{ISM: jets and outflows --- ISM: molecules --- ISM:
abundances}


\section{Introduction}

\begin{figure}
\centering
\includegraphics[angle=0,width=7cm]{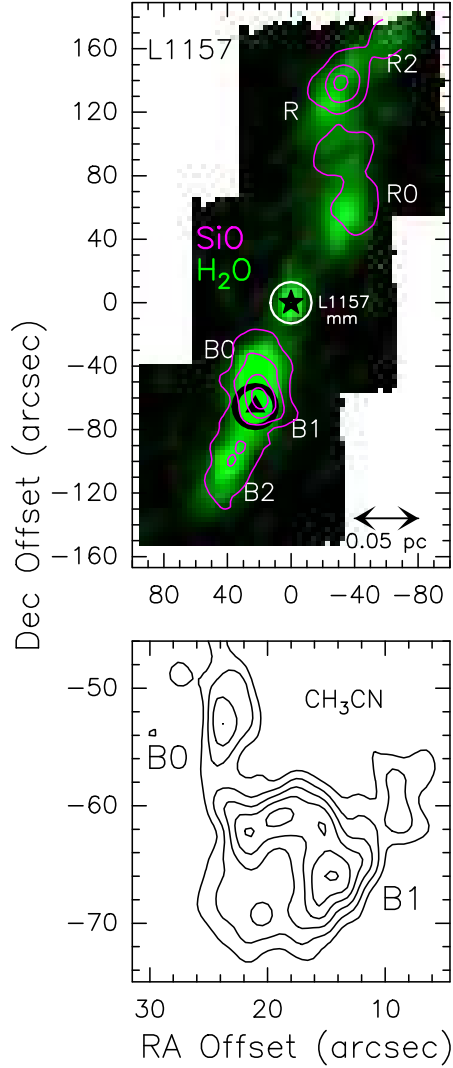}
\caption{{\it Top panel}: PACS image of L1157 of the integrated
H$_2$O emission at 1669 GHz (Nisini et al. 2010b). Offsets are with
respect to the L1157-mm sources (black star), at coordinates:
$\alpha_{\rm J2000}$ = 20$^{\rm h}$ 39$^{\rm m}$ 06$\fs$2,
$\delta_{\rm J2000}$ = +68$\degr$ 02$\arcmin$ 16$\farcs$0.
Magenta contours refer to the SiO(3-2) IRAM 30-m map reported 
by Bachiller et al. (2001).  
The labels indicate the main blue- and red-shifted knots.
Circles are for the IRAM 30-m HPBW at the N$_2$H$^{+}$(1$\--$0) frequency 
(26$\arcsec$), centred at the driving L1157-mm protostar (observed by BP97 and Tobin et al. 2013), 
and at $\Delta\alpha$ = +25$\farcs$6 and $\Delta\delta$ = --63$\farcs$5
from the driving protostar (present observations, see 
black triangles and coordinates reported in Sect. 2).
{\it Bottom panel}: The L1157-B1 bow shock as traced using
the CH$_3$CN(8--7) $K$ = 0,1,2 emission at 3 mm, observed with  
the IRAM PdB interferometer (Codella et al. 2009).}
\label{maps}
\end{figure}

During the first stages of star formation, highly collimated jets from
new born stars influence the physical structure of the hosting
cloud by sweeping up material, compressing and accelerating the
surrounding environment. The propagation of high velocity
outflows generates
shock fronts triggering endothermic chemical reactions and ice grain
mantle sublimation or sputtering.  At a distance of 250 pc (Looney et
al. 2007), the chemically rich L1157 bipolar outflow (Bachiller \& P\'erez Guti\'errez 1997,
hereafter BP97, Bachiller et al. 2001) is an ideal laboratory to observe
the effects of such shocks on the gas chemistry.
L1157 is 
driven by a low-mass ($\sim$ 4 $L_{\odot}$) Class 0 protostar L1157-mm
and it is associated with several blue-shifted (B0, B1, B2) and
red-shifted (R0, R1, R2) shocks at different ages (see Fig.~\ref{maps}--Top panel), and seen in
both CO (Gueth et al. 1996, 1998), and IR H$_2$ (e.g. Neufeld
et al. 1994, Nisini et al. 2010a).  These shocks (see Fig.~\ref{maps}--Bottom panel), when mapped with
interferometers, reveal a clumpy bow structure (e.g. Tafalla \& Bachiller
1995; Benedettini et al. 2007; Codella et al. 2009) at the apex of different molecular
cavities, corresponding to different mass loss episodes (Gueth et al. 1996). 

Both interferometer and single-dish surveys confirm that the L1157 outflow
is well traced by molecules thought to be released off from the dust
mantles such as H$_2$CO, CH$_3$OH, H$_2$O, and NH$_3$
(e.g. Codella et al. 2010, Lefloch et al. 2010, Vasta et al. 2012) as well as by the
refractory grain cores such as SiO (e.g. Nisini et al. 2007; Gusdorf et al. 2008). 
The abundance of these neutral
molecules are enhanced, and the emission  
shows broad wings (up to 20--30 km s$^{-1}$).  On the
contrary, diazenylium (N$_2$H$^{+}$), usually used as tracer of cold prestellar cores 
(e.g. Caselli et al. 2002), shows a completely different behaviour.
Single-dish (IRAM 30-m) and interferometric (IRAM PdB, SMA, CARMA) 
observations indicate
that N$_2$H$^{+}$ traces only the central condensation L1157-mm 
through narrow (0.4--1.0 km s$^{-1}$) emission and it
has not been observed, to date, towards the outflow component (Bachiller et
al. 2001, Chiang et al. 2010, Tobin et al. 2011, 2012, 2013, Yamaguchi et al. 2012).  
The interferometric maps 
show that the narrow N$_2$H$^{+}$ line traces the protostellar envelope 
elongated along a direction perpendicular to the outflow axis 
(i.e. along a hypothetical disk).
However, by analysing their IRAM PdB data, Tobin et al. (2011) concluded that 
although the overall N$_2$H$^{+}$ velocity structure is unaffected by the  
outflow, the morphology of the  
sligthly blue-shifted emission ($\lvert$$v$--$v_{sys}$$\vert$ $\leq$ 0.8 km s$^{-1}$) 
outlines the outflow cavity walls in the
inner 20$\arcsec$--30$\arcsec$ protostellar environment. 
Tobin et al. (2011) proposed that such emission is due either to outflow
entrainment or to a hypothetical shock near the driving protostar.  
The same suggestion is found in the ATCA N$_2$H$^{+}$(1--0) image of the 
protostellar core CG30 by Chen et al. (2008).
On the other hand, J$\o$rgensen et al. (2004) investigated with BIMA the 
protostellar binary NGC1333-IRAS2A-B at 3mm showing that the spatial
distribution of N$_2$H$^{+}$ peaks towards the nearby starless core IRAS2C,  
and is missing in the outflows. 

Therefore, it is still under debate what
role, if any, N$_2$H$^{+}$ is playing in a shocked gas scenario: 
Is the N$_2$H$^{+}$ emission observed by Tobin et al. (2011) and
that marks the cavity opened up by the outflow due to
just enhanced gas column density or really associated with a shock?
Such question is important,
given that N$_2$H$^{+}$ is considered a standard 
molecular tracer of cold
and quiescent prestellar environments (e.g. Tafalla et al. 2006).

In order to uniquely answer these questions 
it is essential to study a region $not$ associated with a protostar,
as the young (2000 years; Gueth et al. 1996), and bright
bow-shock L1157-B1, located at $\sim$ 69$\arcsec$ ($\sim$ 0.1 pc, see Fig.~\ref{maps}) from the protostar. 
As part of the Herschel\footnote{Herschel is an ESA space observatory
  with science instruments provided by European-led principal
  Investigator consortia and with important participation from NASA.}
Key Program
CHESS\footnote{http://www-laog.obs.ujf-grenoble.fr/heberges/chess/}
(Chemical Herschel Surveys of Star forming regions; Ceccarelli et
al. 2010), L1157-B1 is currently being investigated with a  
spectral survey in the $\sim$80$\--$350 GHz interval using the IRAM
30-m telescope (Lefloch et al. in preparation), and in the $\sim$500$\--$2000 GHz range using the
Herschel HIFI instrument (de Graauw et al. 2010).
We present here the first unambiguous detection of N$_2$H$^{+}$ emission
towards a protostellar shock: the observed broad emission has been modeled
using a simple pseudo-time dependent chemical model,
showing how N$_2$H$^{+}$ can be used to shed light on the 
chemical history of the pre-shock gas.
 
\section{Observations and results}

The N$_2$H$^{+}$(1$-$0) line at 93173.76 MHz\footnote{It refers to the
  brightest hyperfine component: $F_{\rm 1}$,$F$ = 2,3--1,2.}  was
observed towards L1157-B1 with the IRAM 30-m telescope at Pico Veleta
(Spain). 
The pointed coordinates were $\alpha_{\rm
  J2000}$ = 20$^{\rm h}$ 39$^{\rm m}$ 10$\fs$2, $\delta_{\rm J2000}$ =
+68$\degr$ 01$\arcmin$ 10$\farcs$5, 
  i.e. at $\Delta\alpha$ = +25$\farcs$6 and $\Delta\delta$ =
  --63$\farcs$5 from the driving protostar. The IRAM  
survey was performed during several runs in 2011 and 2012, using the
broad-band EMIR receivers and the FTS spectrometer in its 200 kHz
resolution mode, corresponding to a velocity resolution of 0.6 km
s$^{-1}$ at 93.2 GHz.  The main-beam efficiency ($\eta_{\rm mb}$) was
0.75, while the HPBW is 26$\arcsec$. 
All the spectra are reported here in units of main beam temperature
(T$_{\rm mb}$).

\begin{figure} 
\centering
\includegraphics[angle=0,width=8cm]{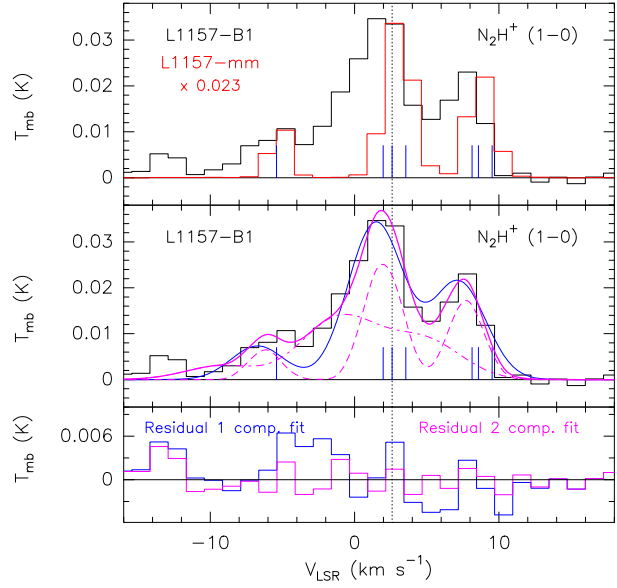}
\caption{{\it Upper panel:} N$_2$H$^{+}$(1$\--$0) line  (black histogram; in T$_{\rm
    mb}$ scale) observed in
L1157-B1 with the IRAM 30-m antenna. The red histogram refers to the N$_2$H$^{+}$(1$\--$0)
spectrum (scaled for a direct comparison) 
as observed towards L1157-mm with the IRAM 30-m antenna in the framework of the ASAI IRAM Large Program (PI: R. Bachiller \&
B. Lefloch). The vertical dashed line
indicates the ambient LSR velocity (+2.6 km s$^{-1}$, from BP97).
The vertical seven blue lines stand for the 15 hyperfine
components of the N$_2$H$^{+}$(1$\--$0) pattern 
(several of them spectrally unresolved at the present
frequency resolution; see Pagani et al. 2009). 
We centered the  spectrum at the frequency of the main hyperfine component $F_{\rm
    1}$,$F$ = 2,3--1,2 (93173.76 MHz). 
{\it Middle panel:} Analysis of the N$_2$H$^{+}$(1$\--$0) profile. The blue line shows the best fit
(FWHM = 4.3 km s$^{-1}$) assuming a single Gaussian component. 
The magenta solid line shows the best fit using two Gaussian components 
(dashed magenta: FWHM = 2.6 km s$^{-1}$; dot-dashed magenta: FWHM = 5.9 km s$^{-1}$) in
order to minimise the residual. The corresponding residuals are reported in the {\it Bottom panel}:
the single component approach gives a 3$\sigma$ (rms = 2 mK) residual.}
\label{n2h+}
\end{figure}

Figure~\ref{n2h+} shows the N$_2$H$^{+}$(1$\--$0) spectrum:   
thanks to the high sensitivity of the IRAM-EMIR receiver
(r.m.s. = 2 mK after smoothing the spectrum to 1.3 km s$^{-1}$), 
we are able to detect the three main groups of
hyperfine components of the $J$ = 1$-$0 transition. 
The integrated intensity is 327$\pm$14 mK km s$^{-1}$.
The N$_2$H$^{+}$ emission in L1157-B1 was hidden in the
noise of the BP97 spectrum, which has 1$\sigma$ rms of 20 mK, definitely larger 
than that of the present dataset (2 mK).

N$_2$H$^{+}$ is a
linear molecular ion in a stable closed-shell $^1$$\Sigma$
configuration. The dominant hyperfine interactions are those between
the molecular electric field gradient and the electric quadrupole
moments of the two nitrogen nuclei (e.g. Caselli et al. 1995),
producing a splitting of the J = 1--0 line into 15 hyperfine
components, characterised by the corresponding quantum numbers $F_{\rm
  1}$ and $F$ (e.g. Pagani et al. 2009).  
To fit the N$_2$H$^+$ spectrum,  
we first assumed a unique velocity component and 
used GILDAS-CLASS90\footnote{http://www.iram.fr/IRAMFR/GILDAS}, 
which gives the best fit (reported in Table 1) of the hyperfine
components (see the blue line in Fig.~\ref{n2h+}--Middle panel). 
The sum of the opacity at the central velocities of all
the hyperfine components $\sum_{\rm i}$$\tau_{\rm i}$ is 0.1$\pm$0.9.
Although the opacity is not well determined the fit indicates
$\sum_{\rm i}$$\tau_{\rm i}$ $\le$ 1, thus suggesting optically thin
emission. Fits fixing $\tau_{\rm i}$ to larger values never gave better results. 

\begin{table}
\caption{Parameters of the hyperfine fits to the N$_2$H$^+$(1--0)$^{a}$ line,
and total column density.}
\label{lines}
\centering
\begin{tabular}{cccccc}
\hline
\multicolumn{1}{c}{$T_{\rm peak}$} &
\multicolumn{1}{c}{rms} &
\multicolumn{1}{c}{$V_{\rm peak}$} &
\multicolumn{1}{c}{$FWHM$} &
\multicolumn{1}{c}{$\sum_{\rm i}$$\tau_{\rm i}$} &
\multicolumn{1}{c}{$N_{\rm tot}$$^c$} \\
\multicolumn{1}{c}{(mK)} &
\multicolumn{1}{c}{(mK)} &
\multicolumn{1}{c}{(km s$^{-1}$)} &
\multicolumn{1}{c}{(km s$^{-1}$)} &
\multicolumn{1}{c}{} &
\multicolumn{1}{c}{(cm$^{-2}$)} \\
\hline
\multicolumn{6}{c}{1 component fit} \\
\hline
34(2) & 2 & +1.3(0.1) & 4.3(0.2) & 0.1(0.9) & 2.4--7.8 10$^{12}$ \\ 
\hline
\multicolumn{6}{c}{2 components fit} \\
\hline
26(2) & 2 & +1.8(0.1) & 2.6(0.1) & 0.2(0.2) & 2.4--8.0 10$^{12}$ \\ 
14(2) & 2 & --1.1(0.4) & 5.9(0.9) & 0.1(0.1) & 0.4--1.3 10$^{12}$ \\         
\hline
\end{tabular}
\begin{center}
$^a$ The spectrum has been centered at the frequency of the
main hyperfine component $F_{\rm 1}$,$F$ = 2,3--1,2 (93173.76). Frequencies have been extracted from the
Cologne Database or Molecular Spectroscopy (M\"uller et al. 2005). See also Pagani et al. (2009).
$^b$ At a spectral resolution of 1.3 km s$^{-1}$.
$^c$ Assuming a T$_{\rm kin}$ = 20-80 K and a source size of 20$\arcsec$--25$\arcsec$ (see text). \\
\end{center}
\end{table}

The peak LSR velocity (+1.3 km s$^{-1}$) of the N$_2$H$^{+}$
profile is sligthly blue-shifted with respect to the ambient velocity (+2.6 km
s$^{-1}$, BP97). The linewidth (4.3 km s$^{-1}$) is also considerably
larger than what observed by BP97 and Tobin et al. (2013) towards the driving protostar
L1157-mm (0.6--0.8 km s$^{-1}$).
This is clearly shown in Figure~\ref{n2h+}, where we report  
the N$_2$H$^{+}$(1$\--$0) line 
(see the red histogram in the Upper panel)
recently observed towards L1157-mm in the framework of the
ASAI\footnote{http://www.oan.es/asai/}
IRAM 30-m Large program (PI: R. Bachiller \& B. Lefloch).
The N$_2$H$^{+}$ profile from the B1 shock 
is definitely broader and more blue-shifted that 
what observed towards the L1157-mm protostar,
indicating a different origin.
Note also that the weak, but not blended,
$F_{\rm 1}$,$F$ = 0,1--1,2 line at $\sim$ --8 km s$^{-1}$ from
the main hyperfine component clearly shows blue-shifted emission.

The best fit of Fig.~\ref{n2h+} shows a
non-negligible residual ($\sim$ 3$\sigma$; see Bottom panel) 
at about --4.0 km s$^{-1}$, which suggests 
non-Gaussian emission from gas at high blue-shifted velocity. 
Indeed a definitely more satisfactory fit can be obtained by assuming two blue-shifted 
Gaussian components 
(see the magenta lines in Fig.~\ref{n2h+} and Table 1): (i) a line centered at +1.8 km s$^{-1}$
with FWHM = 2.6 km s$^{-1}$, plus (ii) a broader (5.9 km s$^{-1}$) line peaking at --1.1 km s$^{-1}$
(dashed and dot-dashed magenta lines in Fig.~\ref{n2h+}, respectively). 
In summary, despite the complexity due to the hyperfine components, 
this clearly shows that a single-Gaussian component is insufficient to reproduce 
the N$_2$H$^{+}$(1--0) profile towards the B1 shock, and one needs to invoke additional
broad blue-shifted emission.
The present observation
thus reports the first detection of N$_2$H$^{+}$ emission towards
a low-mass outflow, definitely far from the protostellar environment.

\section{Physical conditions of the N$_2$H$^+$ gas}

The line profiles in L1157-B1, as in other molecular shock
spots, have a relatively complex structure where several excitation  
components are visible. Disentangling such components is not an easy
task. In L1157-B1, the recent CO multi-line analysis by Lefloch et
al. (2012) indicates that the line profiles are composed by a linear
combination of exponential curves $I_{\rm CO}$($v$) = $I_{\rm CO}$(0)
exp(--$\lvert$$v$/$v_{\rm 0}$$\rvert$), independently of the CO
transition considered. 
The three velocity components correspond to three
different physical components: (1) a small ($\sim$
7$\arcsec$--10$\arcsec$) dissociative J-type shock called $g1$ (identified where
the line intensity is $\propto$ exp(--$\lvert$$v$/12.5$\rvert$))
dominating at the highest velocities ($\le$ --20 km s$^{-1}$), (2) the
outflow cavity walls, $g2$ ($\propto$
exp(--$\lvert$$v$/4.4$\vert$)), with size $\le$ 20$\arcsec$, and (3)
the larger ($\sim$ 25$\arcsec$) outflow cavity created by the older
bow shock L1157-B2, $g3$ ($\propto$
exp(--$\vert$$v$/2.5$\vert$)) dominating at velocities close to the
systemic one ($v$ $\ge$ --2 km s$^{-1}$).  Each component shows the
same slope at all $J$, but different relative intensities. The higher
is the line excitation the brighter is the $g1$ component.  On the
contrary, $g3$ is observed only towards the low--$J$ ($\le$ 3) CO
lines.

\begin{figure}
\centering
\includegraphics[angle=0,width=8cm]{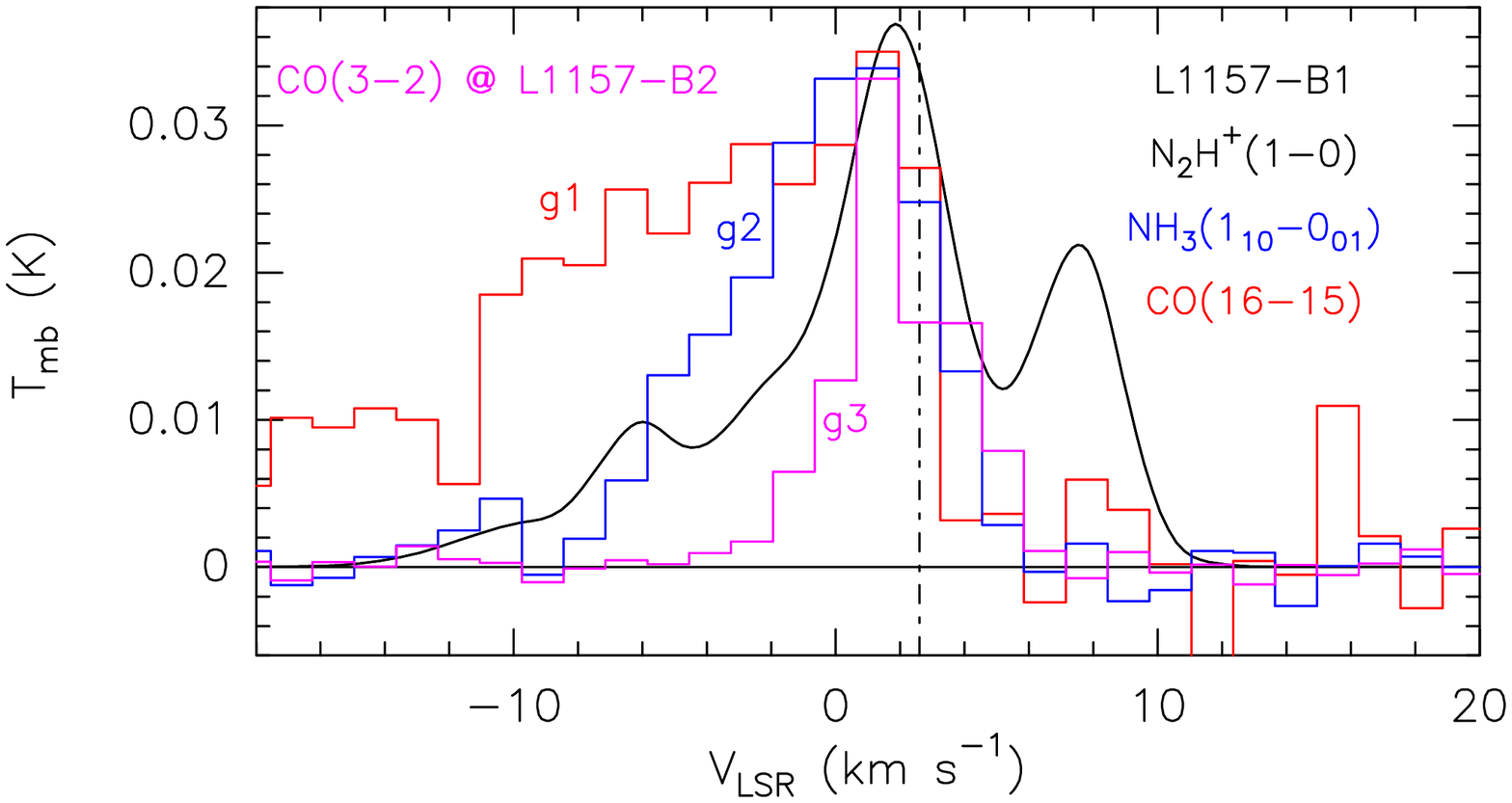}
\caption{Comparison of the N$_2$H$^{+}$(1$-$0) fit (black; see Fig. 2) with typical
  profiles of the $g1$, $g2$, and $g3$ components 
  (from Codella et al. 2010, and Lefloch et al. 2012, see text).
  CO(16--15) represents $g1$ (1841.3 GHz, red, decreased by a factor 6.4 for a
  direct comparison), while NH$_3$(1$_{0}$$-$0$_{0}$) (572.5 GHz, blue,
  decreased by a factor 4.5) is for $g2$. In addition, we  
  report the CO(3--2) spectra for $g3$ (magenta, 345.8 GHz,
  decreased by a factor 269.0) observed by Lefloch et al. (2012)
  towards the L1157-B2 position, tracing a cavity older than the
  L1157-B1 one, and created by a previous wind ejection (Gueth et
  al. 1996).  The spectra have been smoothed to a common spectral resolution
  of 1.3 km s$^{-1}$.}
\label{compa}
\end{figure}

Figure~\ref{compa} compares the
N$_2$H$^{+}$(1--0) line with other line profiles observed towards
L1157-B1 (Lefloch et al. 2010, Codella et al. 2010, 2012): (i)  the
CO(16--15) at 1841.3 GHz observed with Herschel-HIFI as an example of
a spectrum where the $g1$ component is clearly dominating the line
profile; (ii) the CO(3--2) profile, {\it as observed
  towards L1157-B2}, representing a pure $g3$ profile, without the $g1$
and $g2$ components observed towards L1157-B1;  (iii) the
NH$_3$(1$_{0}$$-$0$_{0}$)  
transition, showing a profile well 
reproduced by the $g2$ component alone.  
The N$_2$H$^{+}$ line profile, despite the blending between hyperfine
components, seems to exclude the extremely high-velocity emission
associated with the $g1$ component, being consistent with the $g2$ and $g3$ ones.
In conclusions, N$_2$H$^+$ is
associated either with the B1 outflow cavity (with $T_{\rm kin}$
$\simeq$ 70 K and $n_{\rm H_2}$ $\ge$ 10$^5$ cm$^{-3}$, according
to the LVG CO analysis by Lefloch et al. 2012) and/or with the older and
colder B2 cavity ($\sim$ 20 K, $\ge$ 6 $\times$ 10$^4$ cm$^{-3}$).

The low excitation N$_2$H$^{+}$(1$-$0) transition ($E_{\rm u}$ = 5 K)
has a critical density of $\sim$ 10$^5$ cm$^{-3}$ (e.g. Friesen et al. 2009). 
The line emission is thus expected to be close to LTE conditions at the
densities of the $g2$ and $g3$ gas
components.  Following the results of the LVG analysis by Lefloch et
al. (2012), we assume a $T_{\rm kin}$ between 20 and 70 K and an
emitting size of 20$\arcsec$--25$\arcsec$. The N$_2$H$^{+}$ total
column density is then well constrained $N$(N$_2$H$^+$) = (2--8) $\times$ 10$^{12}$
cm$^{-2}$. 
Using the source-averaged column density $N$(CO) 
= 1 $\times$ 10$^{17}$ cm$^{-2}$ (found for both $g2$ and
$g3$ by Lefloch et al. 2012), and assuming [CO]/[H$_2$]=10$^{-4}$, we
can derive the N$_2$H$^+$ abundance: $X$(N$_2$H$^{+}$) = 2--8 $\times$
10$^{-9}$. A lower abundance, between 4 $\times$ 10$^{-10}$ and $\sim$ 10$^{-9}$, is derived 
for the weaker emission at higher velocity, represented
by the velocity component peaking at --1.1 km s$^{-1}$ (see Table 1).

These values are consistent with what found   
towards the L1157-mm protostar by BP97 (4 $\times$
10$^{-9}$) using the IRAM 30-m antenna. On the other hand, Chiang et al. (2010)    
measured lower values (3--6 $\times$ 10$^{-10}$) towards L1157-mm using the CARMA array,
possibly due to interefometric filtering. 
Similar values have been also found in CO depleted  
prestellar cores and dense protostellar 
envelopes ($\sim$ 10$^{-10}$--10$^{-9}$; see e.g. Caselli et al. 2002,
Tafalla et al. 2004, 2006, Maret et al. 2007, Chen et al. 2007, 2008). 
This value represents an estimate of the abundance of
the gas in the outflow cavities and will be used for a comparison with
the outputs predicted by our models.

\section{N$_2$H$^{+}$ chemistry in L1157-B1}

\begin{figure}
\centering
\includegraphics[angle=0,width=8.5cm]{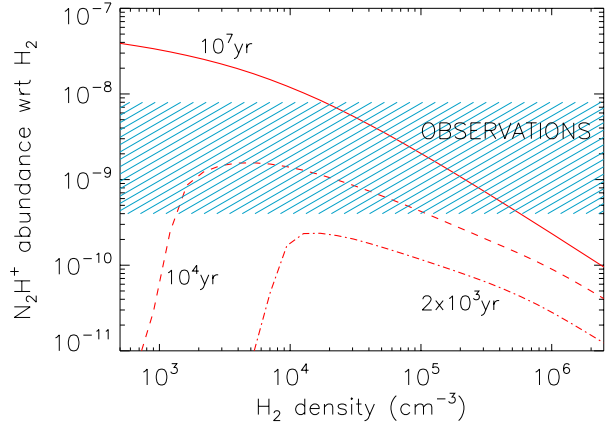}
\caption{N$_2$H$^+$ abundance, with respect to H$_2$, versus the H$_2$ density at different times: 
from $2 \times 10^3$ yr (the age of L1157-B1) to $1 \times 10^7$ yr. The dashed blue box gives the observed value 
with the 1 $\sigma$ uncertainty (see text). 
The gas is at a temperature of 70 K, but the curve is identical in the range 20 to 70 K. The cosmic ionisation rate
is 1$0^{-16}$ s$^{-1}$.} 
\label{model}
\end{figure}

To understand the origin of the observed N$_2$H$^+$, we compared its measured 
abundance with  
the N$_2$H$^+$ abundance predicted by a simple pseudo-time dependent model.  
We used the  
publicy available ASTROCHEM code\footnote{http://smaret.github.com/astrochem/}. The code follows the evolution of the 
chemical composition of a gas cloud initially in the diffuse state and with fixed temperature and density. 
A simple gas-grain interaction due to freeze-out, thermal, and photo-desorption, has been considered.  
In these calculations we assumed a nitrogen elemental
abundance equal to $2.1 \times 10^{-5}$ (with respect to H nuclei), carbon and
oxygen equal to $7.3 \times 10^{-5}$ and $1.8 \times 10^{-4}$ respectively, grain size of 0.1 $\mu$m, 
and cosmic ionisation rates $\zeta$ in the $10^{-17}$--$10^{-16}$ s$^{-1}$ range (e.g. Dalgarno 2006; Padovani et al. 2009). 

Figure~\ref{model} shows the predicted N$_2$H$^+$ abundance as a function of the volume density 
at different evolutionary times, from $2 \times 10^3$ yr (the age of L1157-B1) to $1 \times 10^7$ yr. 
The chemistry of N$_2$H$^+$ is relatively simple: it is formed by the reaction
of the H$_3^+$ (created by the cosmic rate ionisation of H$_2$) and destroyed by the reaction 
of CO (or electrons in case of CO depletion). 
Therefore, the larger the density the lower is the H$_3^+$ abundance, and consequently $X$(N$_2$H$^+$).
The comparison of the measured and predicted N$_2$H$^+$ abundances yields an important conclusion: 
the observed N$_2$H$^+$ abundance is perfectly matched by a model of cold, quiescent, and 
relatively old ($\ge$ 10$^4$ yr) 
gas and does not require the intervent of a shock. The age of the shock in L1157-B1 
is around 2000 yr (Gueth et al. 1996); hence  
Fig.~\ref{model} shows that N$_2$H$^+$ was present before the shock occurred, and it is consistent
with a pre-shock H$_2$ density of $\leq$ 5 $\times$ 10$^{5}$ cm$^{-3}$.
In addition, given that $X$(e) $\propto$ $n_{\rm H_2}$$^{-1/2}$ (e.g. McKee 1989),   
we can $speculate$ that the lower $X$(N$_2$H$^+$) abundance 
(by a factor $\simeq$ 5--6) measured at the highest velocities indicates  
a density gradient in the shocked gas in the cavity. In other words, the N$_2$H$^+$ emitting at higher velocities
could trace gas with $n_{\rm H_2}$ about one order of magnitude higher than that of the gas
at velocities closer to the systemic one.

\begin{figure}
\centering
\includegraphics[angle=0,width=8.5cm]{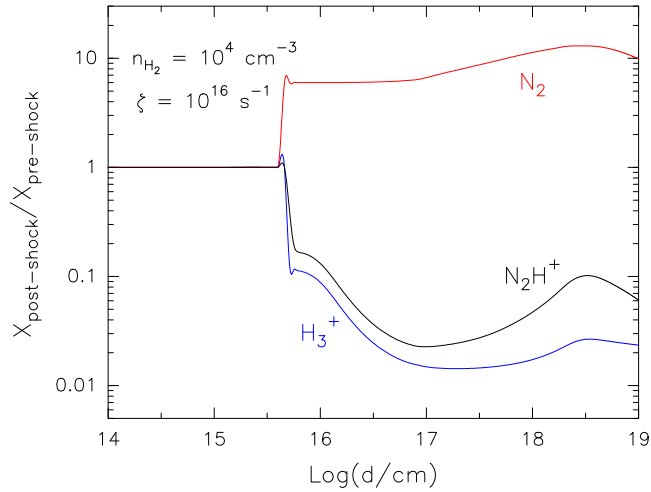}
\caption{Example of UCL\_CHEM model (Viti et al. 2004) showing how the fractional abundances (with respect to H$_2$) 
of N$_2$H$^+$, H$_3$$^+$, and N$_2$ can vary as a function of distance (see text).} 
\label{ucl}
\end{figure}

In addition, to verify whether the detected N$_2$H$^+$ molecules were pre-existing to the shock, 
we used a shock model of L1157-B1 reported by Viti et al. (2011),
who coupled the chemical model UCL\_CHEM with a parametric shock model (Jim\'enez-Serra et al. 2008).
UCL\_CHEM is a gas-grain chemical code which first simulates the formation of high-density clumps from
an atomic diffuse cloud, and then follows their chemical evolution when subjected to the passage of a C-type shock. 
Full details of the code can be found in Viti et al. (2004, 2011).
We updated the grid of models from Viti et al. (2011) varying the cosmic ray ionisation rate $\zeta$ (which of course directly
influences the behavour of ions) in the 10$^{-17}$--10$^{-16}$ s$^{-1}$ range.
Figure~\ref{ucl} reports an example of a UCL\_CHEM shock model
assuming $\zeta$ = 10$^{-16}$ s$^{-1}$ and a pre-shock density of 10$^{4}$ cm$^{-3}$.
We confirm that N$_2$H$^+$ is indeed formed in the gas phase and that the passage of a shock, with the
subsequent release of N$_2$ into the gas, does not yield an increase in the
N$_2$H$^+$ abundance. This is consistent with the lack of signatures
of very high-velocity associated with the $g1$ component in the N$_2$H$^+$(1--0) profile.
On the contrary, the passage of a shock does decrease the N$_2$H$^+$ abundance
by about 1--2 orders of magnitude, depending on the pre-shock conditions and velocity of the shock.
This allows us to further constrain the pre-shock density to $\sim$ 10$^4$ cm$^{-3}$
(see Fig.~\ref{model}) in order to mantain the observed abundance once the outflow cavities have been compressed to $\geq$ 10$^5$ cm$^{-3}$.
A value of $\zeta$ ($\sim$ 10$^{-16}$ s$^{-1}$) helps to achieve and mantain a high N$_2$H$^+$ abundance.
A pre-shock density of $\sim$ 10$^4$ cm$^{-3}$ is consistent with the results suggested by the study of deuteration
in L1157-B1 (Codella et al. 2012b) where it was found that the most likely scenario is that of a
a gas passing through a pre-shock phase with $n_{\rm H_2}$ $\le$ 4 $\times$ 10$^4$ cm$^{-3}$, during which
formaldehyde and methanol ices are formed.

\section{Conclusions}

We present the first detection of diazenylium towards
outflowing gas far from the driven low-mass protostar. 
We found evidence that N$_2$H$^+$(1--0) emission observed towards the L1157-B1 shock  
originates from the dense ($\ge$ 10$^5$ cm$^{-3}$) gas associated
with the cavities
opened, and accelerated by the prototellar wind. 
The line width ($\ge$ 4 km s$^{-1}$) is significantly broader than 
the N$_2$H$^+$ line widths previously observed towards the driving protostar L1157-mm
($\leq$ 1 km s$^{-1}$), as well as than the typical 
line widths observed in quiescent regions, probably as a result 
of the energy injection from the sweeping outflow. 
The estimated N$_2$H$^+$ 
abundance is (2--8) $\times$ 10$^{-9}$, which can be reproduced by a model of quiescent gas 
evolved for more than 10$^4$ yr (i.e. older than the
shock kinematical age, 2000 yr). 
In other words, N$_2$H$^+$ can be considered a fossil record of the pre-shock phase, when the
gas density was $\sim$ 10$^4$ cm$^{-3}$. Modelling of C-shocks  
confirms that $X$(N$_2$H$^+$) is not enhanced by the passage of the shock. 
The present N$_2$H$^+$ detection is the result of the increase of
its column density due to the compression (by a factor $\sim$ 10) of swept-up material, and not to its relative abundance. 

\begin{acknowledgements}
C. Codella, C. Ceccarelli, B. Lefloch, and S. Viti acknowledge the financial support from
the COST Action CM0805 ``The Chemical Cosmos''. 
The Italian authors gratefully acknowledge funding from Italian Space Agency (ASI) through the contract I/005/011/0,
which also supports the fellowships of G. Busquet and A. G\'omez-Ruiz. 
C. Ceccarelli and B. Lefloch acknowledge funding from the French Space Agency CNES and the National Research Agency funded project
FORCOM, ANR-08-BLAN-0225.
S. Viti acknowledges support from the [European Community's] Seventh Framework Programme [FP7/2007-2013] under grant agreement n$^{\circ}$ 238258.
\end{acknowledgements}

\vspace{0.5cm}
\noindent
{\bf References} \\

\noindent
Bachiller R., Per\'ez Guti\'errez M., Kumar M. S. N., et al., 2001, A\&A 372, 899 \\
\noindent
Bachiller R., \& Per\'ez Guti\'errez M., 1999, ApJ 487, L93 (BP97) \\
\noindent
Benedettini M., Viti S., Codella C., et al., 2007, MNRAS 381, 1127  \\
\noindent
Caselli P., Myers P.C., Thaddeus P., 1995, ApJ 455, L77 \\
\noindent
Caselli P., Benson P.J., Myers P. C., Tafalla M., 2002, ApJ 572, 238  \\
\noindent
Ceccarelli C., Bacmann A., Boogert A., et al., 2010, A\&A 521, L22 \\
\noindent
Chen X., Launhardt R., Bourke T.L., Henning Th., Barnes P.J., 2008, ApJ 683, 862 \\
\noindent
Chen X., Launhardt R., Henning Th., 2007, ApJ 669, 1058  \\
\noindent
Chiang H.-F., Looney L.W., Tobin J.J., \& Hartmann L., 2010, ApJ 709, 470  \\
\noindent
Codella C., Benedettini M., Beltr\'an M.T., et al. 2009, A\&A 507, L25 \\
\noindent
Codella C., Lefloch B., Ceccarelli C., et al., 2010, A\&A 518, L112 \\
\noindent
Codella C., Ceccarelli C., Bottinelli S., et al., 2012a, ApJ 744, L164 \\
\noindent
Codella C., Ceccarelli C., Lefloch B., et al., 2012b, ApJ 757, L9 \\
\noindent
Dalgarno A., 2006, Proceedings of the National Academy of Science 103, 12269 \\
\noindent
Friesen R.K., Di Francesco J., Shirley Y.L., Myers P.C., 2009, ApJ 697, 1457 \\
\noindent
de Graauw Th., Helmich F.P., Phillips T.G., et al., 2010, A\&A 518, L6 \\
\noindent
Gusdorf A., Pineau des For$\hat {\rm e}$ts G., Cabrit S., Flower D.R., 2008, A\&A 490, 695 \\
\noindent
Gueth F., Guilloteau S., \& Bachiller R., 1996, A\&A 307, 891 \\
\noindent
Gueth F., Guilloteau S., \& Bachiller R., 1998, A\&A 333, 287 \\
\noindent
Jim\'enez-Serra I., Caselli P., Mart\'{\i}n-Pintado J., Hartquist T.W., 2008, A\&A 482, 549 \\
\noindent
J$\o$rgensen J.K., Hogerheijde M.R., van Dishoeck E.F., Blake G.A., Sch\"{o}ier F.L., 2004, A\&A 413, 993 \\
\noindent
Lefloch B., Cabrit S., Codella C., et al., 2010, A\&A 518, L113  \\
\noindent
Lefloch B., Cabrit S., Busquet G., et al., 2012, ApJ 757, L25  \\
\noindent
Looney L. W., Tobin J., \& Kwon W., 2007, ApJ 670, L131 \\
\noindent
Maret S., Bergin E.A., \& Lada C.J., 2007,  ApJ 670, L25 \\
\noindent
McKee, C.F., 1989, ApJ 345, 782 \\
\noindent
M\"uller H.S.P., Sch\"oier F.L., Stutzki J., Winnewisser G., 2005, J.Mol.Struct. 742, 215 \\
\noindent
Neufeld D.A., \& Green S., 1994, ApJ 432, 158 \\
\noindent
Nisini B., Codella C., Giannini T., et al., 2007, A\&A 462, 163  \\
\noindent
Nisini B., Giannini T., Neufeld D.A., et al., 2010a, ApJ 724, 69  \\
\noindent
Nisini B., Benedettini M., Codella C., et al., 2010b, A\&A 518, L12 \\
\noindent
Padovani M., Galli D., \& Glassgold A.E., 2009, A\&A 501, 619  \\
\noindent
Pagani L., Daniel F., \& Dubernet M.-L., 2009, A\&A 494, 719 \\
\noindent
Tafalla M., \& Bachiller R., 1995, ApJ 443, L37  \\
\noindent
Tafalla M., Myers P.C., Caselli P., Walmsley C.M., 2004, A\&A 416, 191  \\
\noindent
Tafalla M., Santiago-Garc\'{\i}a J., Myers P.C., et al., 2006, A\&A 455, 577  \\
\noindent
Tobin J.J., Hartmann L., Chiang H.-F., et al., 2011, ApJ 740, 45  \\
\noindent
Tobin J.J., Hartmann L., Bergin E., et al., 2012, ApJ 748, 16  \\
\noindent
Tobin J.J., Bergin E., Hartmann L., et al., 2013, ApJ 765, 18 \\
\noindent
Wilson T.L., \& Rood R., 1994, ARA\&A 32, 191 \\
\noindent
Vasta, M., Codella, C., Lorenzani, A., et al., 2012, A\&A 537, A98 \\
\noindent
Viti S., Collongs M.P., Dever J.W., McCoustra M.R.S., Williams D.A., 2004, MNRAS 354, 1141  \\
\noindent
Viti S., Jim\'enez-Serra I., Yates J.A., et al., 2011, ApJ 740, L3
\noindent
Yamaguchi T., Takano S., Watanabe Y., et al., 2012, PASP 64, 105

\end{document}